\documentstyle[psfig, referee]{mn2e}

\def\etal{{\it et al. }}
\title[NGC~4649 Globular Clusters]
{Gemini/GMOS Imaging of Globular Clusters in the Virgo 
Galaxy NGC~4649 (M60)}

\author[Forbes \etal ]{
Duncan A. Forbes$^{1}
\thanks{dforbes@astro.swin.edu.au}$,
Favio Raul Faifer$^{2}
\thanks{favio@fcaglp.fcaglp.unlp.edu.ar}$,
Juan Carlos Forte$^{2}
\thanks{forte@fcaglp.fcaglp.unlp.edu.ar}$,
Terry Bridges$^{3}
\thanks{tjb@astro.queensu.ca}$,\\
\\ \LARGE 
Michael A. Beasley$^{1,4}
\thanks{mbeasley@ucolick.org}$,
Karl Gebhardt$^5
\thanks{gebhardt@astro.as.utexas.edu}$,
David A. Hanes$^3
\thanks{hanes@astro.queensu.ca}$,
Ray Sharples$^6
\thanks{R.M.Sharples@dur.ac.uk}$, \\
\\ \LARGE 
Stephen E. Zepf$^7
\thanks{zepf@pa.msu.edu}$
\\
\\
$^1$ Centre for Astrophysics \& Supercomputing, Swinburne University,
Hawthorn, VIC 3122, Australia\\
$^2$ CONICET and Facultad de Cs. Astronomicas y Geofisicas, UNLP, Paseo del Bosque
       1900, La Plata, Argentina\\
$^3$ Department of Physics, Queen's University, Kingston ON K7L
3N6, Canada\\
$^4$ Lick Observatory, University of California, Santa Cruz, CA 95064, USA\\
$^5$ Astronomy Department, University of Texas, Austin TX 78712, USA\\
$^6$ Department of Physics, University of Durham, South Road,
Durham DH1 3LE, United Kingdom\\
$^7$ Department of Physics and Astronomy, Michigan State
University, East Lansing MI 48824, USA\\
}
\begin{document}
\maketitle

\begin{abstract}

We present Sloan g and i imaging from the GMOS instrument on the Gemini North
telescope for the globular cluster (GC) system around the Virgo
galaxy NGC~4649 (M60). Our three pointings, taken in good seeing
conditions, cover an area of about 90 sq. arcmins. We detect
2,151 unresolved sources. 
Applying colour and
magnitude selection criteria to this source list gives 995
candidate GCs that is greater than 90\% complete to a magnitude
of i = 23.6,  
with little contamination
from background galaxies. 
We find fewer than half a dozen potential Ultra Compact
Dwarf galaxies around NGC 4649. Foreground extinction from the
nearby spiral NGC 4647 is limited to be A$_V$ $<$ 0.1. 
We confirm the bimodality in the GC colour distribution found by
earlier work using HST/WFPC2 imaging. 
As is commonly seen in other galaxies, the red GCs are
concentrated towards the centre of the galaxy, having a steeper
number density profile than the blue GC subpopulation. 
The varying ratio
of red-to-blue GCs with radius can largely explain the overall GC
system colour gradient. 
The underlying galaxy starlight has a similar density
profile slope and colour to the red GCs.
This suggests a direct connection between the galaxy field
stars and the red GC subpopulation.   
We estimate a total GC population of 3700 $\pm$ 900, 
with the uncertainty dominated by the extrapolation to larger
radii than observed. This total number corresponds to a specific
frequency S$_N$ = 4.1 $\pm$ 1.0. 
Future work will present properties derived from GMOS spectra of 
the NGC 4649 GCs.

\end{abstract}

\begin{keywords}
  globular clusters: general -- galaxies: individual: NGC 4649
-- galaxies: star clusters.
\end{keywords}

\section{Introduction}\label{sec_intro}

The study of globular cluster (GC) systems provides important
insights into the star-formation history and chemical enrichment
of galaxies (Ashman \& Zepf 1998, Harris 2001). GCs are 
created during the initial stages of galaxy formation; but they also
appear to form in subsequent star formation episodes up to, and
including, the current epoch.
Observationally, GC systems typically reveal 
a bimodal colour distribution, indicating discrete GC
sub-populations (e.g. Gebhardt \& Kissler-Patig 1999; Larsen \etal
2001; Kundu \& Whitmore 2001)

Several testable galaxy formation scenarios address
the origin of this bimodal behaviour (Ashman \& Zepf 1992,
Forbes, Brodie \& Grillmair 
1997, Cote \etal 1998). Recently, GC formation has been placed
in the broader cosmological context of galaxy formation. 
For example, Beasley \etal (2002)
investigated the formation of GCs in a semi-analytic model which
invoked hierarchical merging in a CDM cosmology, finding that blue
(metal-poor) GCs are formed at high redshift in
proto-galactic fragments, while the red (metal-rich) GCs are
formed later, during the gas-rich merging of these fragments. One
prediction of this model is that the formation of the red GCs
will be extended over several Gyrs in low density environments. 

In order to test the model predictions, and further
probe the halos of galaxies, we have undertaken to study the GC
systems in a sample of early-type galaxies with the Gemini telescopes. The
galaxies cover a range of environments and luminosities. Our
basic approach, using the GMOS instruments, is to obtain deep imaging
of several fields around each galaxy. A subset of the brighter 
candidate GCs are then selected for follow-up multi-object 
spectroscopy (these 
will be presented in a future paper by Bridges \etal
2004).  The spectra allow us to derive 
GC system kinematics plus age and metallicity estimates for
individual GCs. Some initial results of our programme have been
highlighted in Bridges \etal (2003). 

In this paper, we present
and discuss the results from our deep imaging of the GC system
around NGC~4649 (M60). This giant Virgo elliptical 
lies in a subclump to the East of the main Virgo concentration. The
galaxy has strong X-ray emission with a luminosity of L$_X$ = 2.1
$\times$ 10$^{41}$ erg/s
(O'Sullivan, Forbes \& Ponman 2001), largely from a hot diffuse
halo. Chandra imaging reveals the presence of numerous
discrete sources which are identified as low mass X-ray binaries (LMXBs), some of which 
lie in GCs (Sarazin \etal
2003). The galaxy
has an old stellar population with no evidence of a young central
population (Terlevich \& Forbes 2002).

The GC system of NGC~4649 was first studied by Couture, Harris \&
Allwright (1991). Their B and V CCD imaging covered an area of
2.1 $\times$ 3.4 arcmins, and reached a depth of B $\sim$
24.5. They measured a mean colour of B--V = 0.75 for 82 candidate
GCs but could not
detect any bimodality in the colour distribution. A radial colour
gradient was seen in the overall GC population. Harris \etal
(1991) studied the B band GC luminosity function. These data
reached a depth of B $\sim$ 26. They measured the turnover to be
B = 24.47 $\pm$ 0.18 with a Gaussian spread of $\sigma$ =
1.29 $\pm$ 0.12. They also showed that the overall GC system has
a more extended radial distribution than the underlying
starlight. 

Larsen \etal (2001) included NGC~4649 in their HST/WFPC2 study of
GC systems in 17 galaxies, and detected a bimodal GC colour
distribution. 
To V = 25, they detected 176 blue GCs and 169 red
GCs with mean colours of (V--I)$_o$ = 0.954 and (V--I)$_o$ =
1.206. Kundu \& Whitmore (2001) also found bimodality with
peaks around V--I = 0.95 and 1.20. Larsen \etal fit a t$_5$
profile to the GC luminosity function and found a
peak magnitude of V = 23.58 $\pm$ 0.08 and $\sigma$ = 1.28
$\pm$ 0.09. The red peak was 0.2 mag. fainter than the blue
peak. 
The specific frequency in the literature is
relatively high with S$_N$ = 6.7 $\pm $ 1.4 (Ashman \& Zepf 1998). In this
paper we assume a distance to NGC~4649 of m--M = 31.13 (Tonry
\etal 2001) or 16.83 Mpc, which implies 1$^{''}$ = 81 pc. The total
V magnitude from the RC3 is 8.84. The Galactic extinction is  
A$_V$ = 0.086 from Schlegel \etal (1998) and $A_V$ = 0.030 from
Burstein \& Heiles (1982). 
This gives V$_o$ = 8.75 (8.81) which corresponds to an absolute 
magnitude of M$_V$ = --22.38 (--22.32). At a projected distance of 12 kpc
from NGC~4649 lies the SBc spiral NGC~4647.

\section{Observations and Initial Data Reduction}\label{sec_obs}

Images were taken of NGC~4649 using the Gemini Multi-object
Spectrograph (GMOS) in imaging mode on the Gemini North telescope
on 2002 April 10, 11 and 14th. The internal Gemini program ID is 
20020410-GN-2002A-Q-21/Q-13. 
The instrument consists of three 2048 $\times$ 4608
pixel CCDs, with a scale of 0.072 arcsec per pixel, yielding a
5.5 $\times$ 5.5 arcmin field-of-view in a single pointing. We
obtained three pointings close to the galaxy centre: 
the North-East (field No. 1), South-East (field No. 2),  
and East (field No. 3). The
images were taken using modified Sloan g and i filters (centred
at wavelengths of 475 nm and 780 nm respectively). Both g and i
images had exposure times of 4 $\times$ 120 sec, and the 
subexposures were dithered to
remove the gaps between the CCD chips. The seeing for
the three fields were similar in the g and i filters, with 0.6,
0.7 and 1.0 arcsec for fields 1, 2 and 3 respectively.

      The raw images were processed using the Gemini GMOS routines
      within IRAF (e.g. gprepare, gbias, giflat, gireduce
      and gmosaic). The resulting
      images for each filter were then co-added using imcoadd. These
      images were then used for all the subsequent data analysis.

\section{Object Finding and Selection}

\subsection{Galaxy halo removal}

      To aid in the selection, classification and photometry of
      the GC candidates we must first remove the bright background
      of the galaxy halo. 


      After experimenting with different filtering and modelling
      methods we adopted the SExtractor (Bertin \& Arnouts 1996) background
      modelling approach. This method makes a local estimate of
      the background by, first, selecting a number of pixels
      within a box that, after k-sigma clipping and median
      filtering, yields a mesh of background values. These values
      can then be interpolated in order to estimate the
      background level at a given position.
      The best results were obtained for a grid sampling of 32 by
      32 pixels, combined with a 5 pixels (square box) median
      filter, and bi-cubic interpolation.


\subsection{Determination of the Point Spread Function}

         The determination of the point spread function (psf) on
         the co-added images was made using the DAOPHOT II
         package (Stetson 1987) within IRAF. Each psf
         determination involved 30 to 60 objects per field
         and the use of the variable psf option. Some of the 
         unresolved objects (i.e. the brightest and relatively
         uncrowded ones) were used to derive aperture corrections
         to the psf fitting photometry.
             
      An object search was carried out on the i image, since
      this is the deepest in terms of signal-to-noise, using the
      finding routines included in the SExtractor package. This
      approach classifies a detection as real if it is
      composed of a number of connected pixels, above a certain 
signal-to-noise. We used the following SExtractor parameters:
detection treshold of 1.5 sigma; detection minimum area of 3
pixels, and filtering with a Gaussian of FWHM of 0.58 arcsecs.

      A size measure (i.e. resolved or unresolved) for each object was
      made in two steps. The stellarity index, defined in
      SExtractor, was used as a first step. This index takes
      values of around 1.0 for unresolved objects and close to
      0.0 for extended ones (at the distance of NGC~4649, 
      GCs are not resolved in the GMOS images).
 
      Figure 1 shows the behaviour of the stellarity index as a
      function of the i magnitude and shows that, for a range of
      intermediate index values, the nature of a given object
      cannot be unambiguously identified. However from a sample
of 7,000 artifical objects created with the ADDSTARS rountine, we
found that a stellarity index of greater than 0.35 provided a
good initial definition of "unresolved" (i.e. less 1.5\% of
objects were misclassified).

\begin{figure}
\centerline{\psfig{figure=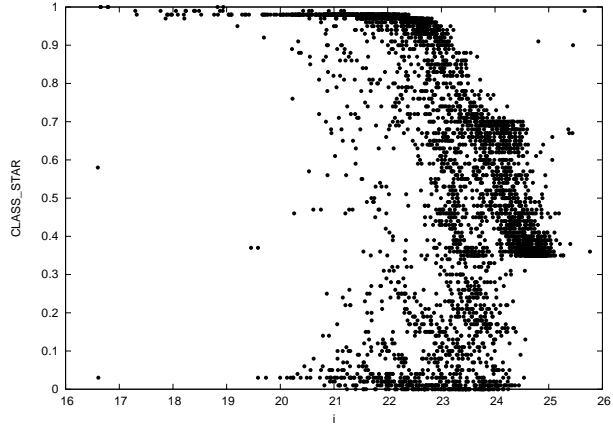,width=0.5\textwidth,angle=270}}
 \caption
{SExtractor stellarity index as a
      function of i magnitude for all the objects found in the
      composite mosaic field. Initially, all objects with an
      index $>$ 0.35 were selected as unresolved.
}
\label{cmd}
\end{figure}

\begin{figure}
\centerline{\psfig{figure=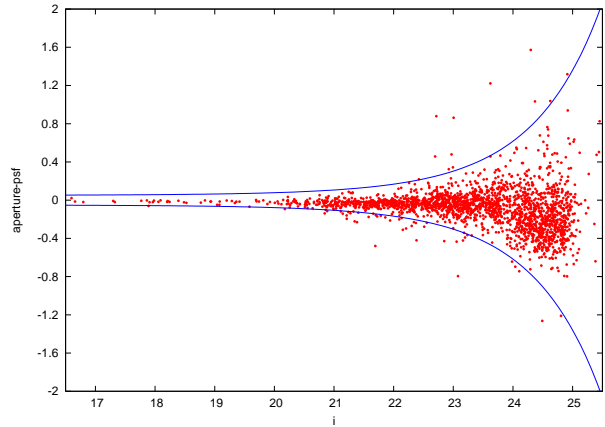,width=0.5\textwidth,angle=270}}
 \caption
{Aperture-minus-psf
      magnitude vs i psf magnitude for all the objects with stellarity index larger
      than 0.35. The solid lines define the upper and lower
      boundaries of the region occupied by objects that were
      finally classified as unresolved.
}
\label{cmd}
\end{figure}

      In order to improve this first selection, we analyzed the position of
      each object on a psf magnitude vs. aperture-minus-psf magnitude
      diagram shown in Figure 2. 
      This figure shows both an increase of the dispersion as a
      function of magnitude, as expected from increasing
      photometric 
      errors, and also an asymmetry towards negative values that
      is indicative of the resolved nature of the images. The
      upper boundary (positive values) in this figure is 
      presumably dominated by unresolved objects and was
      quantified by adopting an exponential fit. This curve, and
      its negative mirror, were then used to
      define the domain of what we consider as unresolved
      objects. All those objects with magnitude differences more
      negative than the lower curve were then removed and added to
      the resolved objects list. Thus, we have a list of 874 
      resolved objects (i.e. galaxies) and a list of 2,151
      unresolved objects, i.e. candidate GCs.  

      Figures 3 and 4 show the spatial distribution of the 
      GC candidates  and of the resolved objects. There is a
      clear concentration of the candidate GCs around the galaxy,
      whereas the resolved objects have a more random spatial
      distribution.

\begin{figure}
\centerline{\psfig{figure=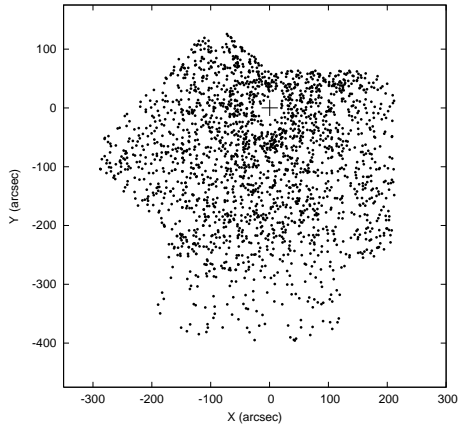,width=0.5\textwidth,angle=270}}
 \caption
{Location of all 2,151 unresolved objects around NGC~4649 in arcsec
  (each pixel is 0.072 arcsec). North is up and East is 
      to the left. There is a clear concentration around the
  galaxy centre (indicated by a cross). 
}
\label{spatial}
\end{figure}

\begin{figure}
\centerline{\psfig{figure=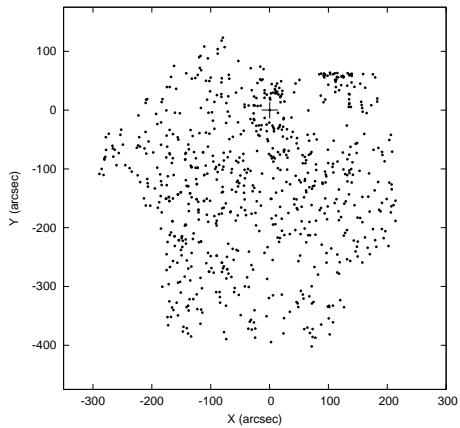,width=0.5\textwidth,angle=270}}
 \caption
{Location of all 874 resolved objects around NGC~4649 in pixels. 
North is up and East to the left. The spatial distribution
appears largely random in nature. 
}
\label{spatial}
\end{figure}

\subsection{Photometry}

      One of the pointings (field No. 1) exhibits the best
      overall seeing both for the g and i images and was 
      adopted as the reference field, i.e. both the instrumental
      (psf) magnitudes and colours of the other two pointings were
      derived using 
      objects in the overlapping regions. After atmospheric
      extinction correction, the instrumental magnitudes and
      colours were linked to the standard system by adopting the
      zero-point values provided by the Gemini pipeline. We
expect the absolute photometric calibration of GMOS images to
improve with time, so our calibration should be considered as
preliminary. The behaviour of the photometric errors for candidate GCs 
(as derived from the DAOPHOT
       package) are shown in Figures 5 and 6. Typical errors at i
      = 24 are 0.08 in magnitude and 0.13 in colour. 

Note that the magnitudes and colours presented in this paper have {\it not}
been corrected for Galactic extinction. The extinction values
from Schlegel \etal (1998; A$_V$ = 0.086) and from
Burstein \& Heiles (1982; $A_V$ = 0.030) with A$_g$ = 1.25A$_V$
and A$_i$ = 0.65A$_V$, imply a colour reddening correction of
0.052 and 0.018 respectively. If we apply the g,i to V,I 
transformations of Smith \etal (2002) 
our colours are $\sim$0.05 mag redder than the
Schlegel \etal extinction-corrected values given in Larsen \etal
(2001). This suggests that our magnitudes, although preliminary,
are close to the standard photometric system.
       
\begin{figure}
\centerline{\psfig{figure=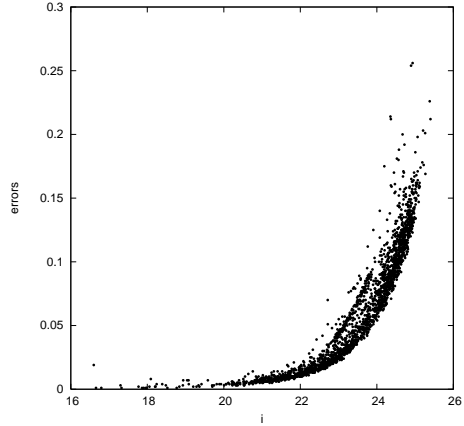,width=0.5\textwidth,angle=270}}
 \caption
{The i magnitude errors as a function of magnitude for the 
       globular cluster candidates listed in Table 1. Typical 
errors are $\pm$ 0.08 mag. at i = 24.
}
\label{cmd}
\end{figure}

\begin{figure}
\centerline{\psfig{figure=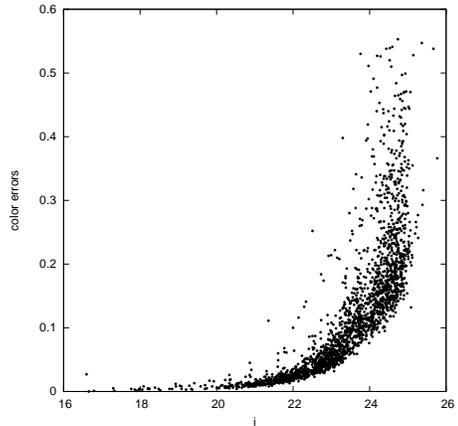,width=0.5\textwidth,angle=270}}
 \caption
{The (g-i) colour errors as a function of the i magnitude
       for the globular cluster candidates listed in Table
1. Typical colour errors are $\pm$ 0.13 mag. at i = 24.
}
\label{cmd}
\end{figure}

      Table 1 gives the photometry for the candidate GCs
      (similar measurements for the resolved objects are given in
      the Appendix Table A1). We list x,y position (in pixels),
      defined in a rectangular system with its origin at the
      galaxy centre with x positive towards West and y positive towards
      the North; galactocentric radius (in
      arcsecs), i magnitude, g--i colour and their errors.  


\begin{table}
\begin{center}
\renewcommand{\arraystretch}{1.0}
\begin{tabular}{lcccccc}
\multicolumn{7}{c}{{\bf Table 1.} Candidate NGC~4649 Globular Clusters}\\
\hline
X & Y & R$_{GC}$ & i & g--i & i err & g--i err \\
(pix) & (pix) & (arcsec) & (mag) & (mag) & (mag) & (mag) \\
\hline
3479  & 4546  &  101.6 & 22.95 &  1.11 &  0.03 &  0.06 \\
4284  & 4546  & 151.3  & 21.19 &  0.99 &  0.01 & 0.01\\
4048  & 4541  & 135.8  & 20.51 & 0.67  & 0.01 & 0.01\\
... &  ... & ... & ... & ... & ... & ... \\
\hline

\end{tabular}
\end{center}
Notes: the full table is available in the on-line version.
\end{table}

\subsection{Completeness experiments}

       An estimate of the completeness of the photometric sample
       given in Table 1 was carried out by means of ADDSTARS
       experiments conducted on the i band galaxy-subtracted image. A total
       number of 7000 artificial objects were added 
to the image and the identification routine
       was run using the same search parameters as before. 
The spatial distribution of
       the artificial objects was determined using a 
       stochastic-generating program that, in an approximate
       way, follows the radial distribution of the globular cluster
       candidates.

The results of our completeness tests are shown in Figures 7 and
8. Figure 7 shows the overall magnitude completeness,
indicating that at i = 23.4 the data are 90\% complete and at i = 24.7
the data are 50\% complete. However, our completeness varies with
galactocentric radius.
       In Figure 8 we show the completeness fractions for several
       different ranges of magnitude as a function
       of galactocentric radius. This figure shows that for i
magnitudes brighter than 23.4, we have detected the vast majority
of GCs over the entire NGC~4649 field. For magnitudes fainter than i
= 23.4, we are missing some of the innermost GCs. 

\begin{figure}
\centerline{\psfig{figure=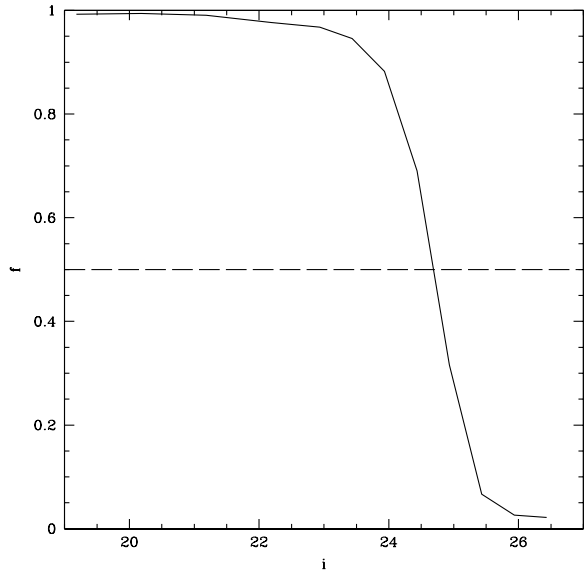,width=0.5\textwidth,angle=0}}
 \caption
{Overall completeness fraction as a function of i magnitude. The
horizontal line shows the 50\% completeness level corresponding
to a magnitude of i $\sim$ 24.7.  
}
\label{cmd}
\end{figure}

\begin{figure}
\centerline{\psfig{figure=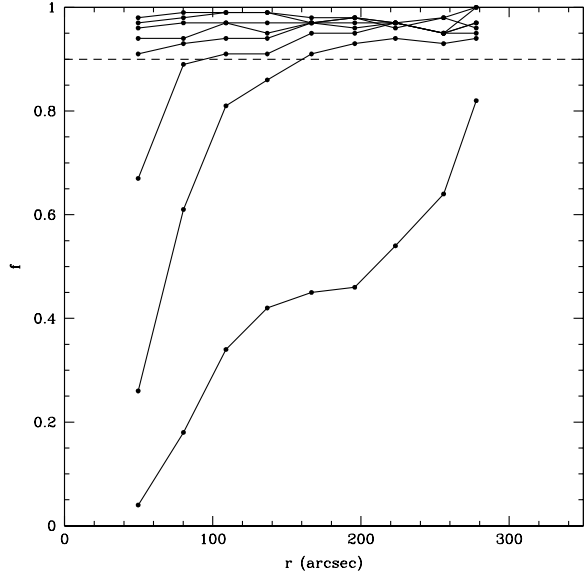,width=0.5\textwidth,angle=0}}
 \caption
{Completeness fraction as a function of
       galactocentric radius for the entire NGC~4649 field, after
galaxy halo subtraction. The
curves show the completeness (from top to bottom) for i = 19.14,
20.14, 21.14, 22.14, 22.9, 23.4, 23.9, 24.4. The horizontal
line shows the 90\% completeness level. 
}
\label{cmd}
\end{figure}

\subsection{Background contamination}


       The degree of background contamination for the objects
       listed in Table 1 was estimated by adopting a comparison
       field, and assuming it is representative of the background
near NGC~4649. This field was observed as part of our overall 
Gemini program, and is located some distance from the galaxy
NGC~7332. Objects in the comparison field were selected using a
similar procedure to that for the NGC~4649 fields. The number of
resolved objects (i.e. presumably background galaxies) 
in the comparison field 
was found to be 256. This is consistent with the 874 resolved 
objects found in the three galaxy fields given the the slight
seeing variations between fields. 
We note also, that we find 38 non-resolved objects 
(i.e. presumably foreground stars) in
the comparison field. The galaxy model of Ratnatunga \& 
Bahcall (1985) predicts 22 in our magnitude and colour range. 

After
completeness tests, similar to those described above, we find
that the
background field has a similar magnitude completeness to the
NGC~4649 data with the
50\% completeness fraction occurring at i $\sim$ 24.7. 
From this comparison field, we estimate the contamination of background
objects in our sample of candidate GCs to be $\sim$3\% for i $<$ 24. 
Thus we can be confident that the effects of background
objects on the results presented below are minimal. 



\section{Foreground dust from NGC~4647 ?}

The spiral galaxy NGC~4647 (UGC~7896) is located to the NW (PA
$\sim$ --45$^{o}$) of NGC~4649 at a projected separation of about
12 kpc. It has a recession velocity of 1422 km/s, or 305 km/s
greater than that of 
NGC~4649. There are no obvious tidal interaction signatures
between the two galaxies. Images of NGC~4647 reveal a flocculent
spiral structure with strong dust lanes. If NGC~4647 lies in the
foreground then dust in its outer regions may cause reddening of the
NGC~4649 GCs in the NW quadrant.

We have compared both the GCs and the underlying starlight in the
NW quadrant to each of the other three quadrants. We find similar
colours ($\Delta$g--i $<$ 0.1) and similar mean magnitudes
($\Delta$i $<$ 0.05). These limit the extinction to be A$_V$ $<$
0.1. In a study of dust extinction in overlapping
galaxies, White, Keel \& Conselice (2000) concluded that if
NGC~4647 caused any additional extinction in NGC~4649 then it was
limited to A$_B$ $<$ 0.11. Thus with no evidence for strong foreground dust 
reddening due to NGC~4647, we 
proceed to treat all GCs the same.

\section{Results}

\subsection{Globular Cluster Colours}

A colour-magnitude diagram for the candidate GCs is shown in
Figure 9. The diagram shows some evidence of colour bimodality,
with colours of g--i $\sim$ 0.9 and 1.2 down to magnitudes of i
$\sim$ 24; fainter than that a broad colour distribution is
seen. The contribution of contaminating background objects 
rises steeply at magnitudes fainter than i = 23.6. 
This magnitude also corresponds to reasonable errors of $\pm$0.05 in i
magnitude and $\pm$0.08 in colour.  Thus we now only consider objects
brighter than i = 23.6. Our bright magnitude limit is chosen to
be i = 20 (M$_I$ = --11.12) as the few sources brighter than this are
likely to be foreground stars or compact dwarf galaxies. We note
in passing that the number of {\it potential} Ultra Compact
Dwarfs (UCDs; Drinkwater \etal 2003) with --12.12 $< $ M$_I$ $<$
--11.12 and g--i $\sim$ 1 is about half a dozen. All of these 
could be foreground stars (see Ratnatunga and Bahcall 1985) 
suggesting that
NGC 4649 does not contain a large population of UCDs. 
A generous colour range of 0.5 to 1.5 has also been applied for
GC selection. Such a selection would include all Milky Way GCs
after a reddening correction. These
magnitude and colour constraints are shown in Figure 9 and used
for the subsequent analysis. A total of 995 objects meet these
selection criteria. 

\begin{figure}
\centerline{\psfig{figure=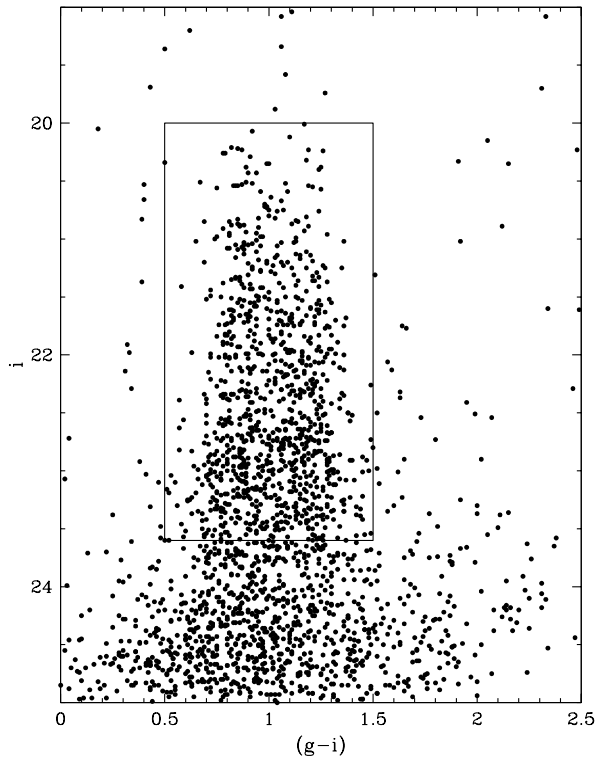,width=0.5\textwidth,angle=0}}
 \caption
{Colour-magnitude diagram for candidate GCs. 
There is some evidence of a bimodal colour
distribution at bright magnitudes. The box shows our selected
colour and magnitude region for subsequent analysis.   
}
\label{cmd}
\end{figure}

In Figure 10 we show the colour distribution, smoothed on 0.05 mag
scales, for the
candidate GCs from the CMD selected region. Two Gaussians have
been fit (using NGAUSS in IRAF) to the distribution allowing their amplitudes, central
position and FWHM to vary. We find that the distribution is well
fit by two Gaussians. The blue subpopulation has a peak color of
g--i = 0.865 $\pm$ 0.005, and the 
red
subpopulation has g--i = 1.167 $\pm$ 0.004 (an extinction
correction would make these colours 0.02--0.05 bluer).

In Figure 11 we show colour distributions, smoothed on 0.05 mag scales, of candidate
GCs over three radial bins (45-90, 90-150 and 150-240
arcsec), all of which have a completeness of $\sim$ 90\% or better. These 
figures show how
the inner regions are dominated by red (metal-rich) GCs, the
outer regions by blue (metal-poor) GCs and intermediate regions
have similar numbers of the two GC subpopulations. 

\begin{figure}
\centerline{\psfig{figure=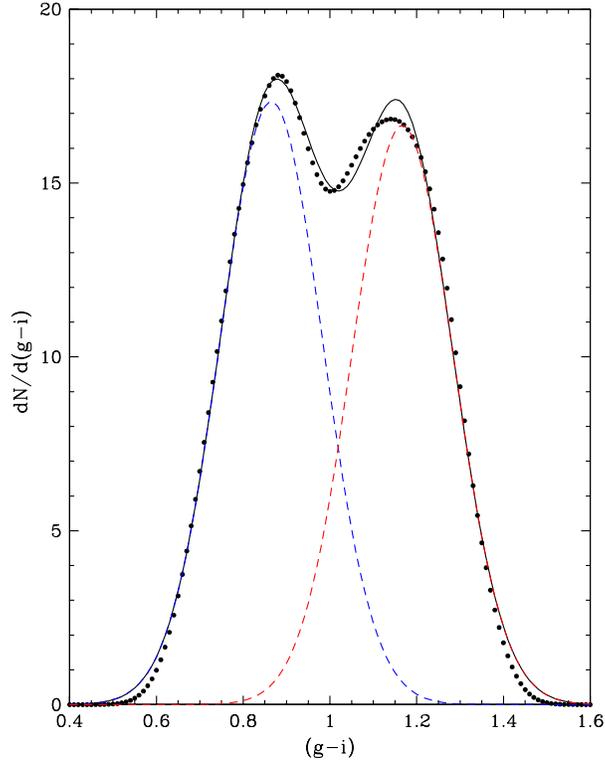,width=0.5\textwidth,angle=0}}
 \caption
{Colour distribution of the magnitude-limited sample of
candidate GCs. The data, smoothed by 0.05 mag., 
are shown by small filled circles. 
The dashed lines represent a Gaussian fit to the
blue and red subpopulations, with the solid line showing the
combined fit. Peaks are found at g--i = 0.87 and 1.17.
}
\label{cmd}
\end{figure}

\begin{figure}
\centerline{\vbox{
\psfig{figure=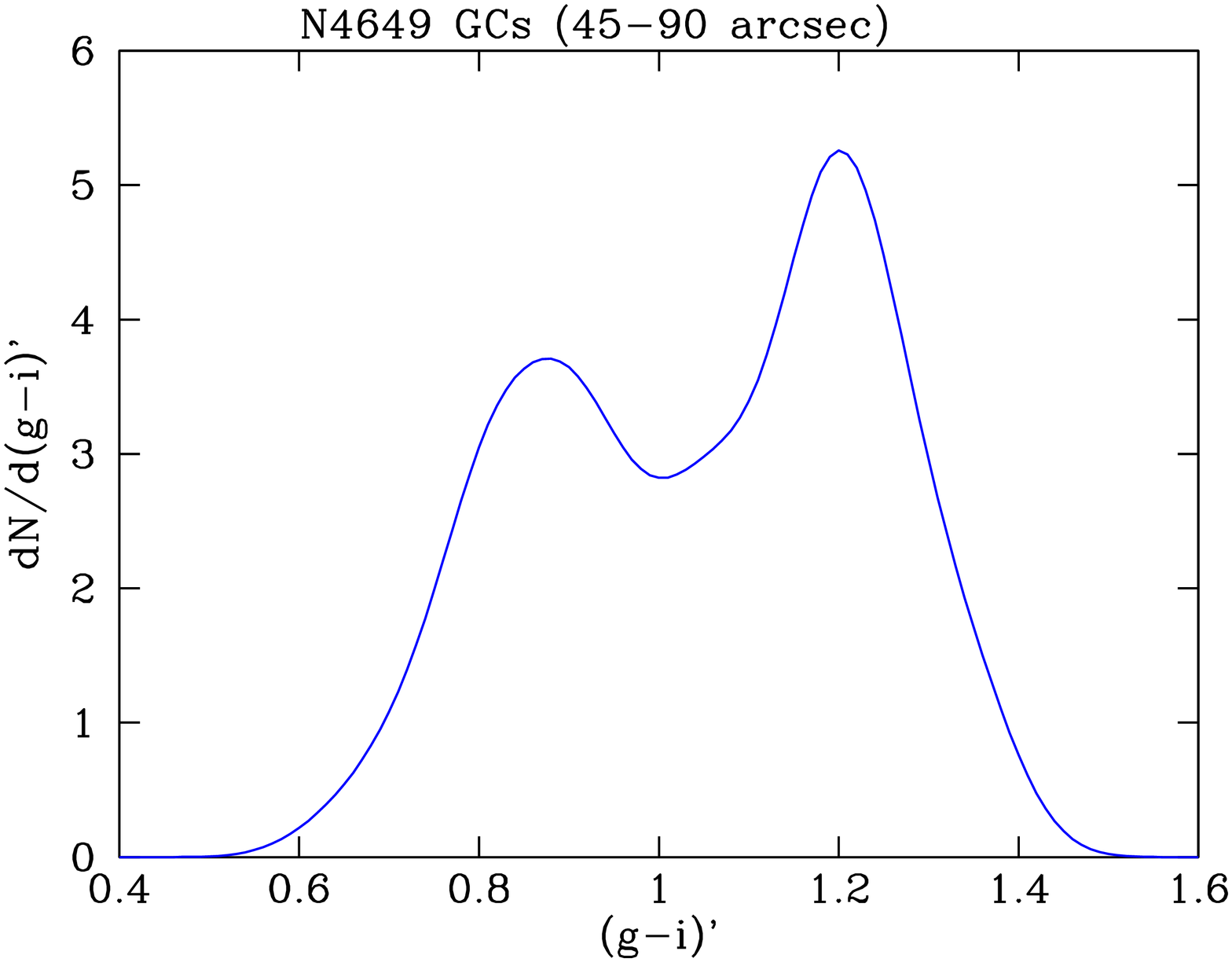,width=0.5\textwidth,angle=270}
\psfig{figure=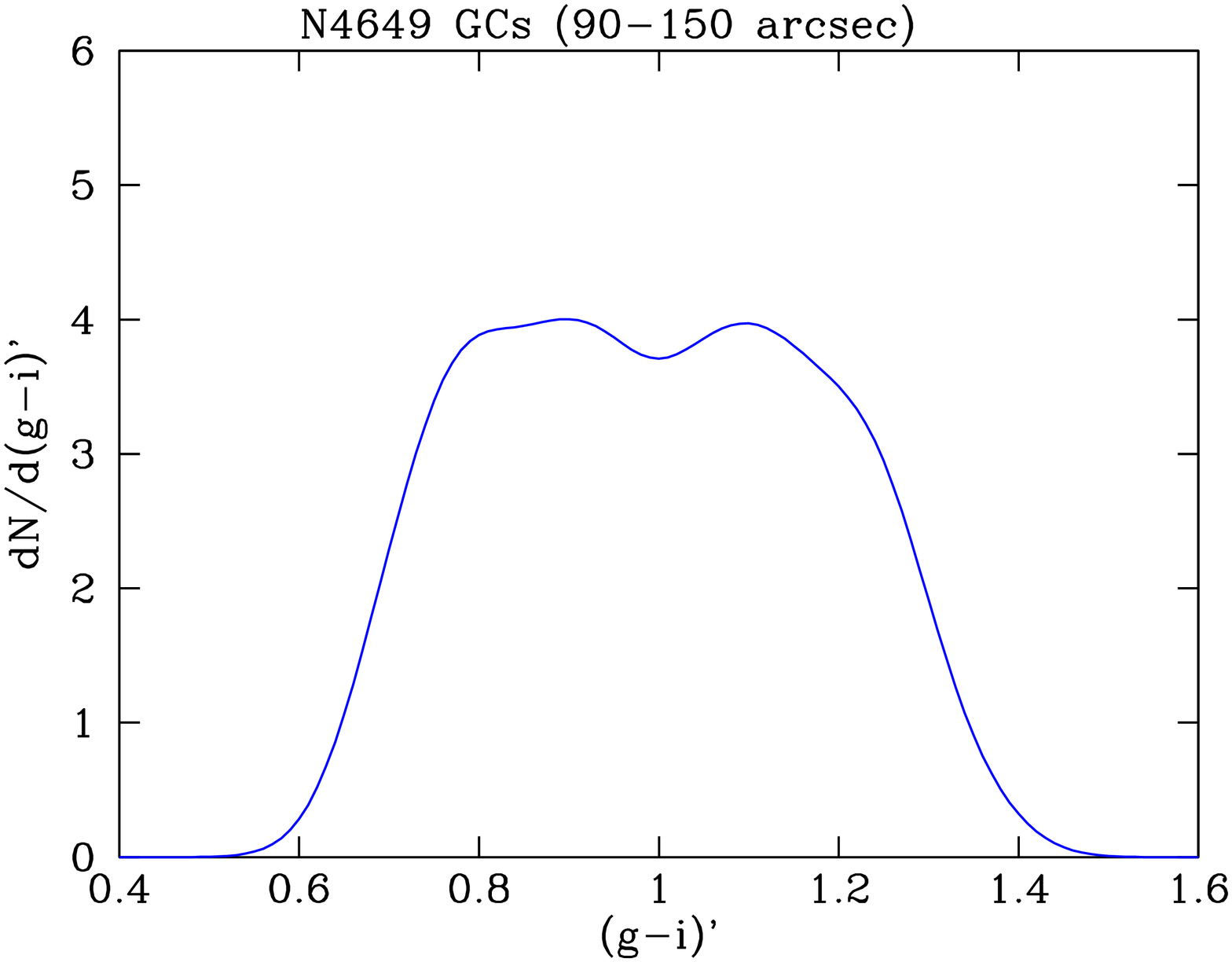,width=0.5\textwidth,angle=270}
\psfig{figure=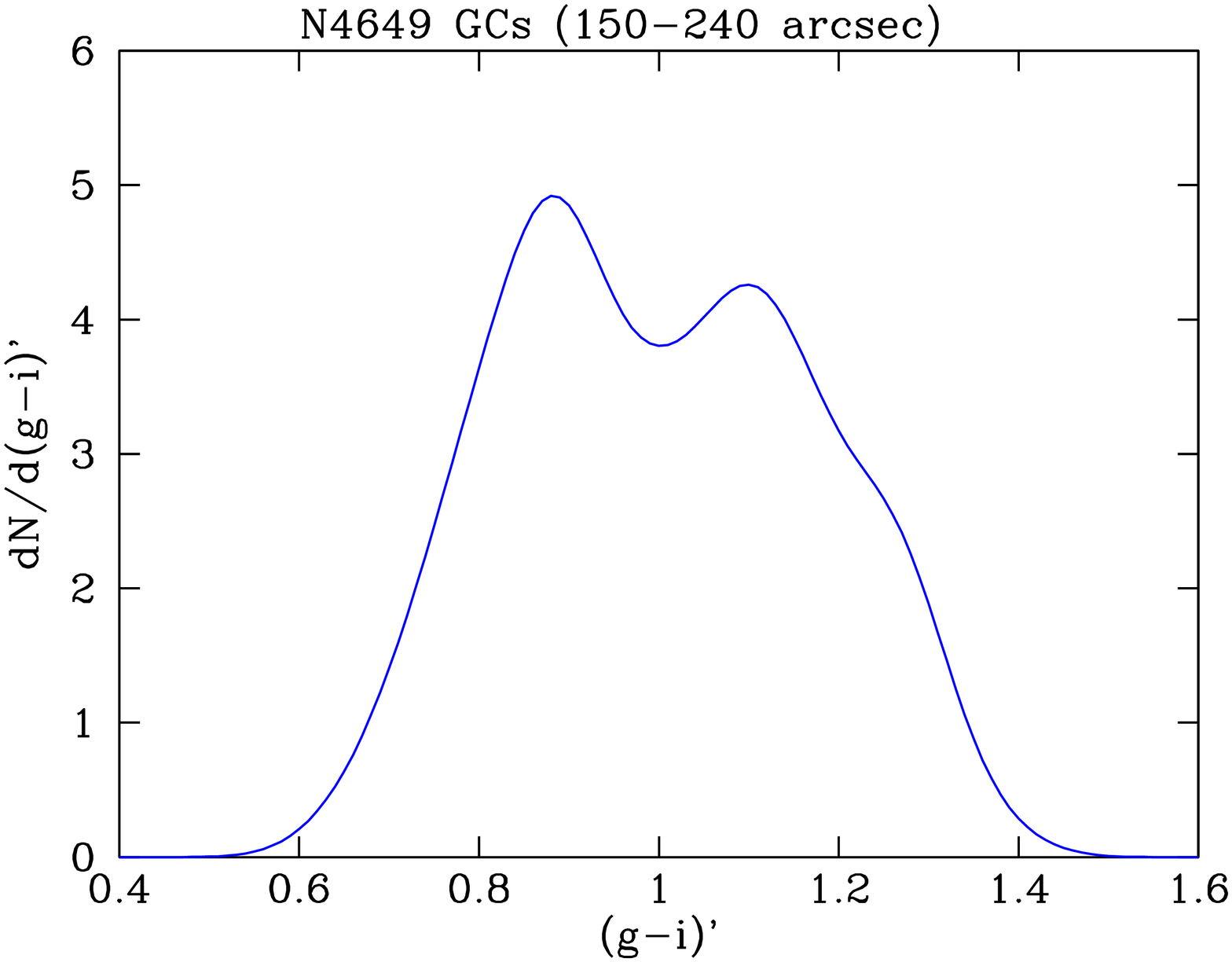,width=0.5\textwidth,angle=270}
}}
\caption{Colour distributions of the 
candidate GCs. The figure shows three radial bins, smoothed by
0.05 mag: {\bf (Top):} 45-90 arcsec; 
{\bf Middle:} 90-150 arcsec; {\bf (Bottom)}:150-240 arcsec. The
red GCs dominate the inner regions. }
\end{figure}

In Figure 12 we show the radial gradient of the GC subpopulations
compared to the overall GC system and the underlying galaxy
starlight. For the galaxy profile two different sky subtraction
algorithms have been used; for one the sky level has been subtracted 
using the counts in the outer regions of our
field-of-view, and for the second these counts have been reduced
by $\sim$75\% (under the assumption that the outer regions still
contains galaxy light). 
Both the red
and blue GC subpopulations are statistically consistent with no radial colour
trend (having non-zero slopes of 2$\sigma$ and 1.5$\sigma$
respectively). However, the 
overall GC population has a radial colour trend of 6$\sigma$
significance, becoming bluer with radius. 
This is
therefore the effect of the changing relative mix of GCs with
radius, i.e. the ratio of red-to-blue GCs decreases with
radius. Similar trends 
been found over the radial range observed here for 
NGC 4472 (Geisler, Lee \& Kim 1996; Rhode \& Zepf 2001) and NGC
1399 (Forbes \etal 1998; Dirsch \etal 2003).  
The galaxy light has a similar colour to the
mean colour of the red GCs for common radii (see also Forbes \&
Forte 2001).

\begin{figure}
\psfig{figure=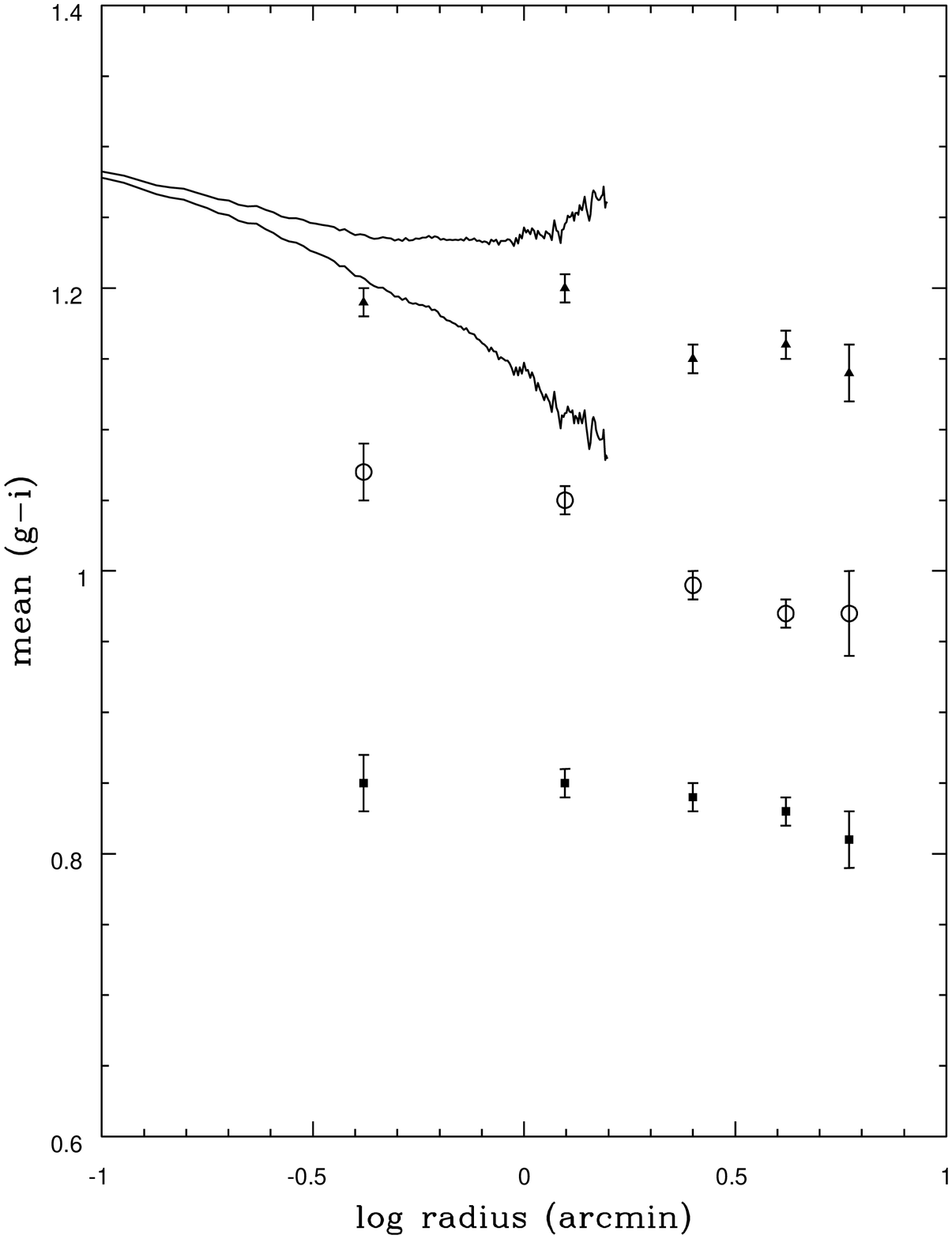,width=0.5\textwidth,angle=0}
\caption{Globular cluster mean colour gradients. 
The red (triangles) and blue (squares)  
GC subpopulations are shown along with the 
total GC population (circles). 
Poisson error bars are given. The solid lines show the g--i
colour profile of the underlying galaxy starlight for two
different sky subtractions. 
Both the red
and blue GC subpopulations are consistent with no radial trend,
whereas the overall GC population becomes bluer with radius. 
The galaxy light is more similar in colour to the
red GCs than the blue subpopulation. 
}
\end{figure}

\subsection{Globular Cluster Surface Density}

In Figure 13 we show the surface density of the two GC
subpopulations versus galactocentric radius. The 
subpopulations have been restricted in colour to 0.5 $< $ g--i $<$
0.9 and 1.1 $<$ g--i $<$ 1.5 to help remove the GCs with
intermediate colours. A correction has been made for the missing
area in each radial annulus. The error bars correspond to Poisson
statistics. 

As expected from
Figure 11, we find the red GCs are more concentrated towards the
galaxy centre than the blue GCs. In other words the red GCs have
a steeper surface density profile. A simple power law fit to the
surface density 
profiles gives a slope of --1.73 $\pm$ 0.06 for the reds and
--1.04 $\pm$ 0.09 for the
blues. We note that Harris \etal (1991) measured a slope of
--1.08 $\pm$ 0.10 for the overall GC system out to radii of
140$^{''}$ (for comparison our overall slope measured out to
230$^{''}$ is --1.3 $\pm$ 0.05). 

Figure 13 also shows the galaxy i band surface brightness
profile, after converting into log units and applying an arbitrary
vertical normalisation. 
The slope of the galaxy profile changes from --1.45 $\pm$ 0.02 in
the inner regions (5$^{''}$ to 30$^{''}$) to --1.75 $\pm$ 0.02 in
the outer regions (30$^{''}$ to 230$^{''}$). 
Thus for the region
of overlap with the GC system, the galaxy starlight has the same
slope, within the errors, to that of the red GC subpopulation.

\begin{figure}
\psfig{figure=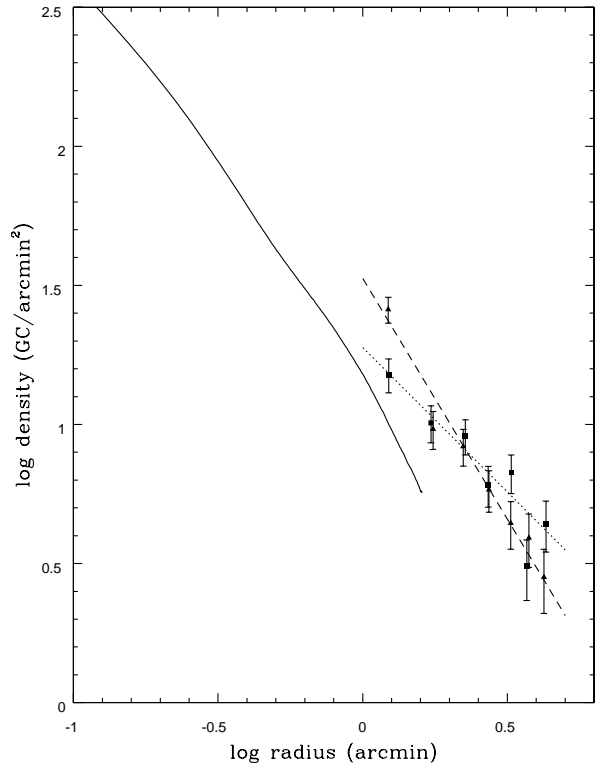,width=0.5\textwidth,angle=0}
\caption{Globular cluster surface density profiles. The red (triangles)
and blue (squares) GC subpopulations are shown
separately. Poisson error bars are given. The dashed line shows
the best fit to the red GCs with power slope of --1.73, and the
dotted line the blue fit of --1.04. The solid line shows the
galaxy starlight profile, which has a slope consistent with the
red GCs. }
\end{figure}

\subsection{GC Specific Frequency}

Given our somewhat uncertain photometric calibration we will not
attempt a detailed analysis of the GC luminosity function for
NGC~4649. However we can still usefully estimate the total number
of GCs in the NGC~4649 system, and hence the GC specific frequency.

We have corrected the counts, in 0.25 mag bins, for our
incompleteness as a function of galactocentric radius and
magnitude. As mentioned in section 3.4, for i $<$ 23.6 we have
essentially detected all GCs over the field-of-view of our
observations. We need to make two corrections in order to estimate the
total number of GCs in the system. The first concerns the missing
areal coverage, i.e. beyond 230$^{''}$, and the second concerns
GCs fainter than the limiting magnitude of i = 23.6. 

By integrating the surface density profile from 60$^{''}$ to
230$^{''}$ we estimate 1007 GCs. Interior to 60$^{''}$ we assume
a constant GC number density as adopted by Harris \etal
(1991). This gives another 219 GCs. Extrapolating the surface
density profile out to 10 arcmins or 48.6 kpc (such large radial
extents have been observed for giant ellipticals in clusters)
gives an additional 1421 GCs. 

\begin{figure}
\psfig{figure=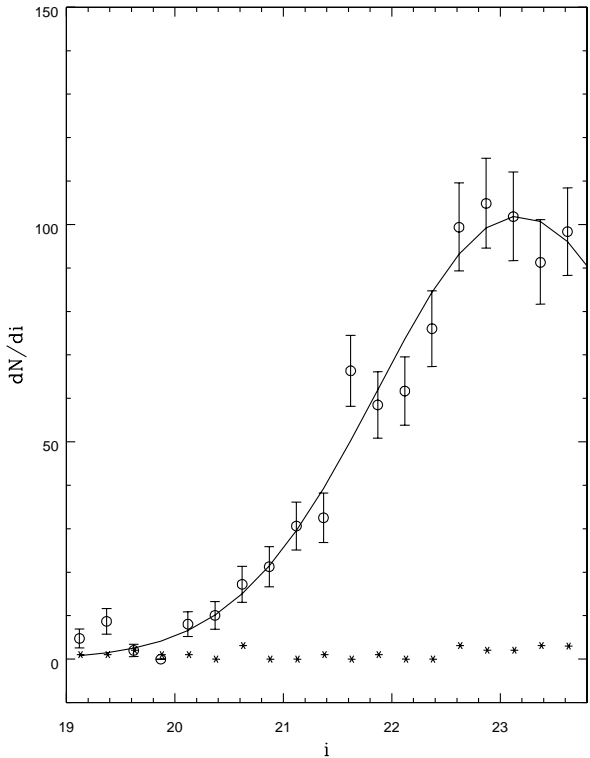,width=0.5\textwidth,angle=0}
\caption{Globular cluster luminosity function. The data, in 0.25
mag bins, are shown by open circles with Poisson error
bars. The data have been corrected for incompleteness and 
background contamination (the background level is shown by stars).
The solid line shows a Gaussian fit to the
data for magnitudes i $<$ 23.6. The fit has a peak or turnover magnitude
at i = 23.17, and a dispersion $\sigma$ = 1.30.
}
\end{figure}

Thus we estimate the total number of GCs spatially to be 
2647 but this is only
to magnitudes of i = 23.6.  
In Figure 14 we show the completeness and background corrected GC luminosity
function in 0.25 mag bins.  The corrected counts 
show a steady Gaussian-like rise profile.
We have fit the data using NGAUSS in IRAF and find a Gaussian profile of width
$\sigma$ = 1.30 $\pm$ 0.1 mag (Harris
\etal 1991 found $\sigma$ = 1.29 and Larsen \etal 2001 measured
$\sigma$ = 1.28) and a peak or turnover
magnitude of i = 23.17 $\pm$ 0.15. The resulting Gaussian profile is shown in
Figure 14 to be a reasonable fit to the data. This profile has
about 62\% of GCs brighter than i = 23.6 (the limit used in the
density analysis) and 38\% fainter. 

We estimate the total number of GCs in the NGC~4649 system,
integrated over all magnitudes and area, to be 3653. Various random errors
are present in this estimate, but the largest uncertainty is
probably associated with the extrapolation of
the surface density to large radii not covered by our images.
For example, if we had chosen 8 or 12
arcmins, instead of 10 arcmins, for the radial extent our estimate
would change by $\pm$ 25\%. In addition, the true surface
density profile might not be a single power-law at all
radii (Rhode \& Zepf 2004; Dirsch \etal2003)
A remaining systematic error is that of the 
photometric calibration.
These caveats aside, our final estimate of the total GC population 
is 3700 $\pm$ 900 which is 
less than the 5100 $\pm$
1100 given in the compilation of Ashman \& Zepf (1998).
This value is based on observations by Harris \etal (1991).
We derive a steeper density profile than Harris \etal, and hence
a smaller GC population. 

For an absolute magnitude of M$_V$ = --22.38, we estimate a
global specific frequency of S$_N$ = 4.1 $\pm$ 1.0. This is fairly
typical of cluster ellipticals, and less than the 6.7 $\pm$ 1.4
in Ashman \& Zepf (1998). We also note that because the
ratio of red-to-blue GCs decreases with radius (see section 5.1), the {\it local}
S$_N$ value for red GCs will decrease relative to the blue S$_N$
with galactocentric radius.



\section{Discussion and Conclusions}\label{sec_conc}

Our two filter imaging of the NGC 4649 GC system has returned a
number of results. However, before turning to GCs, we confirm the findings of White \etal (2000)
that any possible foreground extinction due to the disk of NGC
4647 is limited to be less than $\sim$0.1 mag. 
We also
argue that the number of Ultra Compact Dwarfs associated with NGC
4649 is small, i.e. less than half a dozen. 

With the caveat that the GMOS photometric calibration is not
final, we detect GC colour bimodality with subpopulations at g--i
= 0.865 $\pm$ 0.005 and 1.167 $\pm$ 0.004 (an extinction
correction would make these colours 0.02--0.05 bluer). 

As is commonly seen in other galaxies, the red GCs are
concentrated towards the centre of the galaxy. They have a steep
number density profile of slope --1.73 $\pm$ 0.06 compared to the
shallower slope of --1.04 $\pm$ 0.09 for the blues. 
The varying ratio
of red-to-blue GCs with radius can largely explain the overall GC
system colour gradient. 

For common radii, 
the underlying galaxy starlight has a similar density
profile to the red GCs and colour to the mean colour of the 
red GCs. This 
suggests similar metallicities and/or ages (formation epoch) 
for the red GCs and the galaxy field
stars. A luminosity-weighted spectral age based on Lick indices (Terlevich \& Forbes
2002) suggests that the galaxy stars are old. 
The fact that the peak of the red GC luminosity function is
fainter than that for the blue GCs (Larsen \etal 2001) is consistent with an old
age for the GCs, as at a given age metal-rich GCs are expected to
be fainter than metal-poor GCs due to line blanketing effects.
Our forthcoming GMOS spectra will have
sufficient S/N to derive Lick style ages for the brighter GCs
(Bridges \etal 2004). 

We estimate a total GC population of 3700 $\pm$ 900, which
corresponds to a specific frequency S$_N$ = 4.1 $\pm$ 1.0. This
is lower than the current value quoted in the literature.

\section{Acknowledgments}\label{sec_ack}

We thank the Gemini support staff for their help. We thank
R. Musgrave for help preparing some of the figures, and M. Pierce
for his comments on the text. We also thank the referee,
S. Larsen, for suggested improvements to the paper. These data
were based on observations obtained at the Gemini Observatory,
which is operated by the Association of Universities for Research
in Astronomy, Inc., under a cooperative agreement with the NSF on
behalf of the Gemini partnership: the National Science Foundation
(United States), the Particle Physics and Astronomy Research
Council (United Kingdom), the National Research Council (Canada),
CONICYT (Chile), the Australian Research Council (Australia),
CNPq (Brazil), and CONICET (Argentina). The Gemini program ID 
is 20020410-GN-2002A-Q-21/Q-13. 
This work was supported
in part by a research grant provided to DAH by the Natural
Science and Engineering Research Council of Canada. 

\section{Appendix A}

The photometric data for the resolved objects in the field of NGC
      4649 are given in Table A1.  The magnitudes are not total magnitudes but derived from 
      aperture photometry. The colours however should be largely
unaffected. 

\begin{table}
\begin{center}
\renewcommand{\arraystretch}{1.0}
\begin{tabular}{lcccccc}
\multicolumn{7}{c}{{\bf Table A1.} Resolved Objects in the field
of NGC~4649}\\
\hline
X & Y & R$_{GC}$ & i & g--i & i err & g--i err \\
(pix) & (pix) & (arcsec) & (mag) & (mag) & (mag) & (mag) \\
\hline
3925 & 4531 & 127.7 & 21.81 & -0.30 & 0.03 & 0.04\\
3812 & 4545 & 121.1 & 24.24 & -0.35 & 0.10 & 0.12\\
3818 & 4511 & 120.3 & 21.32 & -0.02 & 0.03 & 0.04\\
... &  ... & ... & ... & ... & ... & ... \\
\hline

\end{tabular}
\end{center}
Notes: the full table is available in the on-line version.
\end{table}

\end{document}